\documentclass[10pt,letterpaper,twocolumn]{article}

\usepackage[margin=2.54cm]{geometry}
\usepackage{color}
\usepackage{lineno}
\usepackage{amsmath}
\usepackage{amssymb}
\usepackage{graphicx}
\usepackage[round,comma,authoryear]{natbib}
\bibpunct[]{(}{)}{,}{a}{}{,}
\usepackage{float}
\usepackage{setspace}
\usepackage{comment}
\usepackage{booktabs}
\usepackage{lscape}
\usepackage{bm}
\usepackage{titlesec}
\usepackage[labelsep=period, figurename=Fig.\ ]{caption}

\titleformat*{\section}{\normalsize\bfseries}
\titleformat*{\subsection}{\normalsize\bfseries}

\title{
A multistate dynamic site occupancy model for \\ spatially aggregated sessile communities
}
\author{
  {\small Keiichi Fukaya,$^{1,6}$ J. Andrew Royle,$^{2}$ Takehiro Okuda,$^{3}$
  Masahiro Nakaoka,$^{4}$ and Takashi Noda$^{5}$} \\\\
  {\footnotesize \it $^{1}$The Institute of Statistical Mathematics, 10-3 Midoricho, Tachikawa, Tokyo 190-8562 Japan} \\
  {\footnotesize \it $^{2}$USGS Patuxent Wildlife Research Center, 12100 Beech Forest Road, Laurel, Maryland 20708 USA} \\
  {\footnotesize \it $^{3}$National Research Institute of Far Seas Fisheries, Japanese Fisheries Research and Education Agency,} \\
  {\footnotesize \it 2-12-4 Fukuura, Kanazawa-ku, Yokohama, Kanagawa, 236-8648 Japan} \\
  {\footnotesize \it $^{4}$Akkeshi Marine Station, Field Science Center for Northern Biosphere, Hokkaido University,} \\
  {\footnotesize \it Aikappu, Akkeshi, Hokkaido 088-1113, Japan} \\
  {\footnotesize \it $^{5}$Faculty of Environmental Earth Science, Hokkaido University,} \\
  {\footnotesize \it N10W5, Kita-ku, Sapporo, Hokkaido 060-0810, Japan} \\
  {\footnotesize $^{6}$kfukaya@ism.ac.jp}
}

\date{ }

\begin{document}

\twocolumn[
  \begin{@twocolumnfalse}
    \maketitle

    \begin{abstract}
    Inferring transition probabilities among ecological states
    is fundamental to community ecology because,
    based on Markovian dynamics models, 
    these probabilities provide quantitative predictions about community composition
    and estimates of various properties that characterize
    community dynamics.
    Markov community models have been applied to sessile organisms
    because such models facilitate estimation of transition probabilities
    by tracking species occupancy at many fixed observation points over
    multiple periods of time.
    Estimation of transition probabilities of sessile communities
    seems easy in principle but may still be difficult in practice because 
    resampling error (i.e., a failure to resample exactly the same location at
    fixed points) may cause significant estimation bias.
    Previous studies have developed novel analytical methods to correct for
    this estimation bias. However, they did not consider the local structure of
    community composition induced by the aggregated distribution of organisms
    that is typically observed in
    sessile assemblages and is very likely to affect observations.
    In this study, we developed a multistate dynamic site occupancy model
    to estimate transition probabilities
    that accounts for resampling errors associated with local community structure.
    The model applies a nonparametric multivariate kernel smoothing methodology
    to the latent occupancy component to estimate
    the local state composition near each observation point,
    which is assumed to determine the probability distribution of data
    conditional on the occurrence of resampling error.
    By using computer simulations, we confirmed that 
    an observation process that depends on local community structure
    may bias inferences about transition probabilities.
    By applying the proposed model to a real dataset of
    intertidal sessile communities, we also showed that estimates of transition probabilities
    and of the properties of community dynamics may differ considerably 
    when spatial dependence is taken into account.  
    Our approach can even accommodate an anisotropic spatial correlation of species composition,
    and may serve as a basis for inferring complex nonlinear ecological dynamics.\\

    \end{abstract}
  \end{@twocolumnfalse}
  ]


\section*{INTRODUCTION}

Markov models are a general class of mathematical models that are widely used
to describe dynamics of ecological systems.
In the context of community ecology, they offer a simple and useful representation of
community dynamics.
For example, a very basic Markov model summarizes the dynamics of a community (i.e.,
changes in community composition or site occupancy dynamics among
ecological states) with a transition matrix, $\mathbf{P}$, which consists of the
transition probability from state $k$ to $j$, $p_{jk}$, in the element in its $j$ th row and
$k$ th column.
Such linear Markov models may not provide a fully realistic 
description of community dynamics in nature, 
which may be inherently nonlinear \citep{Spencer2008,Tanner2009}.
Nevertheless, they can still provide a good approximation of observed community dynamics
and composition \citep{Wootton2001, Wootton2001E, Hill2002}.
Linear Markov models also allow us to assess a wide range of properties of
species-level dynamics (e.g., rate of colonization, disturbance, and replacement)
as well as those of community-level dynamics
(e.g., equilibrium community composition, as well as its sensitivities,
mean turn over time, and damping ratio), which can be derived from the
transition probability matrix \citep{Hill2004}.

Markov community models have been applied to communities of sessile organisms
(e.g., terrestrial plants, corals and intertidal/subtidal communities)
because in those systems transition probabilities can be estimated with field data
by tracking species occupancy at many fixed observation points over multiple periods of time
\citep[e.g.,][]{Wootton2001,Wootton2001E,Hill2002,Hill2004,Tanner2009}.
However, estimation of transition probabilities may be biased when
there are resampling errors, that is, failures to resample exactly
the same fixed-point locations.
When such a resampling error occurs, researchers observe a point that is
different from, but probably close to, the correct location of the fixed point;
the result may be a ``misclassification'' of the occupancy state.
Such an error may sometimes be inevitable in field sampling, where somewhat crude tools
are necessarily used and/or organisms are small \citep{CCD2011}.
New analytical methods have been proposed to correct this estimation bias which 
assume that the probability distribution of data, conditional on the
occurrence of resampling error, is related to the relative state frequency
of ecological states \citep{CCD2011,Fukaya2013}.
On the one hand, \cite{CCD2011} proposed a maximum likelihood approach,
in which transition probabilities and resampling error rates are
estimated based on multinomial likelihoods.
On the other hand, \cite{Fukaya2013} proposed to use the dynamic site occupancy
modeling framework to account for resampling error in the estimation of 
transition probabilities.

In both of these studies, spatial aspects of community structure
were not considered explicitly. 
It is very likely, however, that the local structure of community composition
affects the observation process described above,
because sessile organisms typically represent locally aggregated spatial distributions.
Therefore, accounting for such spatial dependence will be necessary to
make more reliable inferences about underlying community dynamics.
In this study, we developed an extension of the multistate dynamic site occupancy model
proposed by \cite{Fukaya2013} to account for spatial dependence in
the estimation of transition probabilities.
We used additional information about the spatial coordinate of each observation point
to estimate the state composition \textit{near} each point.
Estimation of this local community structure
was possible by applying the nonparametric multivariate kernel
smoothing methodology \citep{Diggle2005}
to the latent occupancy component.
Our approach can even account for an anisotropic spatial correlation of species composition
and may also lead to some potential extensions of the model,
which we describe in the {\it Discussion}.
By using computer simulations, we confirmed that 
an observation process that depends on local community structure
may cause some bias in the inference of transition probabilities.
By applying the proposed model to a real dataset of
intertidal sessile communities, we also showed that estimates of transition probabilities
and of community dynamics may be considerably different
when spatial dependence is accounted for.

\section*{THE MODEL}

\subsection*{Model description}

We assumed that there were, in total, $I$ fixed observation points within a permanent quadrat
that could be occupied by one of the possible $S$ ecological states (species or groups of species,
which may include ``free space'').
Hence, observation points and the quadrat corresponded to
``sites'' (local populations) and the ``meta-population'', respectively,
in the meta-population design concept for the site occupancy modeling framework
\citep{Kery2012,Kery2015}.
The ecological state was assessed for each site over $T$ periods of time.
We define $z_{it}$
and $y_{itn}$ ($z_{it}, y_{itn} \in 1, \dots, S$) as the 
occupancy state at site $i$ ($i = 1, \dots, I$) at time $t$ ($t = 1, \dots, T$),
and the state observed by a researcher (i.e., data recorded) at the $n$th measurement 
($n = 1, \dots, N(i,t)$) at site $i$ at time $t$, respectively.
The $N(i,t)$ observations consisted of replicated samples at site $i$ and time $t$;
these samples may have been collected by repeated surveys conducted within 
a sufficiently short period or by independent observers.
This sampling scheme formally resembles Pollock's robust design,
which is classically considered in capture-recapture studies
\citep{Pollock1982} in which secondary resampling is performed
within a single primary sampling period ($t$).
We note that the dynamic site occupancy modeling framework described below
also permits missing data.

Changes in site occupancy states ($z_{it}$) across primary sampling periods
are described with a transition probability matrix
$\mathbf{P}$.
The element in the $j$ th row and $k$ th column $p_{jk} (1 \leq j, k \leq S)$
represents the transition probability from state $k$ to $j$.
Resampling error probability $e$ is also considered to account for 
a certain type of observation error that arises when there is failure to resample
the exact location of the fixed observation site \citep{CCD2011}.
It is thus assumed that when resampling error does not occur 
at a sampling occasion for site $i$ at time $t$ (occurring with a probability $1 - e$),
the occupancy state $z_{it}$ is recorded with probability 1,
whereas when resampling error occurs,
the ``occupancy state'' observed is a random variable
that follows a certain probably distribution \citep{Fukaya2013}.
\cite{Fukaya2013} assumed that when a resampling error occurs,
the probability of state $s$ being observed is equal to the relative dominance
of that state within the quadrat,
$f_{ts} = \sum_i \bm{1}(z_{it}=s)$, where $\bm{1}(x)$ is an indicator function.
Conceptually, this assumption implicitly requires that organisms are homogenously distributed
within the quadrat so that the state composition is identical over all sites.
The same assumption was made by \cite{CCD2011} to account for
a similar type of resampling error in longitudinal observations of sessile communities.
To account for the spatial dependence of community structure,
we here consider the relative dominance of the state \textit{near} each site, $g_{its}$,
and relate it, instead of $f_{ts}$, to the data distribution conditional on the
occurrence of the resampling error.

Although the model framework could accommodate more ecological realities, 
we here restricted our model by using the simplest elements possible
for notational simplicity and clarity.
We believe this approach will facilitate understanding the basic model structure.
Possible extensions of the model are described in the {\it Discussion}.
Formally, the model we propose is as follows.

{\bf Observation model ---}
We assume that on each sampling occasion, with probability $(1-e)$,
observers assess the exact location of site $i$
and find the true occupancy state $z_{it}$.
But otherwise (i.e., with probability $e$), they fail to observe
the exact location of that site. In the latter case, we assume that 
they record state $s$ with a probability that equals the local 
dominance of that state, $g_{its}$.
This observation distribution, conditional on the occurrence of the
resampling error, represents a categorical distribution
with a probability vector $(g_{it1}, \dots, g_{itS})$.
With a latent indicator variable $m_{itn}$ representing
the occurrence of resampling error for the $n$th measurement at
site $i$ at time $t$, the model for this observation process is expressed as:
%
\begin{equation}
    \begin{aligned}
        y_{itn} &= z_{it}
                    &       & \text{when $m_{itn} = 0$} \\
        y_{itn} &\sim \textrm{Categorical}(g_{it1}, \dots, g_{itS})
                    & \quad & \text{when $m_{itn} = 1$}.
    \end{aligned}
    \label{observation_2}
\end{equation}

\noindent{We} assume that resampling errors occur independently,
and thus that $m_{itn}$ simply follows a Bernoulli distribution:
%
\begin{equation}
    m_{itn} \sim \textrm{Bernoulli}(e).
    \label{error}
\end{equation}

{\bf Process model ---}
Assuming that the dynamics of the occupancy state is a Markov process,
the conditional probability of latent state $z_{it}$
for $t = 2, 3, \dots, T$ can be expressed as follows:
%
\begin{equation}
    z_{it} |(z_{i,t-1}\hspace{-0.25em}=\hspace{-0.25em}s) \sim
        \textrm{Categorical}(p_{1s}, p_{2s}, \dots, p_{Ss}),
    \label{process}
\end{equation}

{\noindent 
where} $p_{js}$ gives the transition probability from state $s$ to $j$.
{\noindent 
The} transition probability matrix $\mathbf{P}$ is expressed as follows:
%
\begin{equation}
    \mathbf{P} = \begin{pmatrix}
        p_{11} & \dots & p_{1s} & \dots & p_{1S} \\
        \vdots & \ddots & \vdots & & \vdots \\
        p_{s1} & \dots & p_{ss} & \dots & p_{sS} \\
        \vdots & & \vdots & \ddots & \vdots \\
        p_{S1} & \dots & p_{Ss} & \dots & p_{SS}
    \end{pmatrix}
    \label{trans.mat}
\end{equation}

{\noindent
where} column $s$ gives the vector of transition probabilities
for $z_{it}|(z_{i,t-1}\hspace{-0.25em}=\hspace{-0.25em}s$).
Note that this transition probability matrix is 
column-stochastic (i.e., each column sums to 1).

For $t = 1$, the initial occupancy probability of each site is
defined by a probability vector
$\boldsymbol{\phi} = (\phi_{1},\dots,\phi_{S})$. 
The model for initial occupancy states is thus expressed as:
%
\begin{equation}
    z_{i1} \sim
        \textrm{Categorical}(\phi_{1}, \dots, \phi_{S}).
    \label{inits}
\end{equation}

{\bf A kernel estimator for the local state composition ---}
The local state dominance $g_{its}$ is unobserved, and we now describe a model 
for $g_{its}$.
We estimate $g_{its}$ nonparametrically by using a multivariate kernel 
smoothing methodology to estimate the spatial structure of 
multinomial probabilities \citep{Diggle2005}.

We let $\mathbf{x}_i$ be a two-dimensional coordinate vector for site $i$.
For each $i$, $t$, and $s$, the kernel regression estimator for 
the local state composition is expressed as follows:
%
\begin{equation}
  g_{its} = \frac{\sum_{j = 1}^I K_{\boldsymbol{\Sigma}}(\mathbf{x}_i, \mathbf{x}_j)\bm{1}(z_{jt} = s)}{\sum_{r = 1}^S \sum_{j = 1}^I K_{\boldsymbol{\Sigma}}(\mathbf{x}_i, \mathbf{x}_j)\bm{1}(z_{jt} = r)},
  \label{kernel.est2}
\end{equation}

\noindent{where} $z_{jt}$ is the latent occupancy state at site $j$ at time $t$, and 
$K_{\boldsymbol{\Sigma}}(\mathbf{x}_i, \mathbf{x}_j)$ is a
two-dimensional Gaussian kernel function with a bandwidth matrix $\mathbf{\Sigma}$:
%
\begin{equation}
  K_{\boldsymbol{\Sigma}}(\mathbf{x}_i, \mathbf{x}_j) = \exp \left\{ -\frac{1}{2} (\mathbf{x}_i - \mathbf{x}_j)' \mathbf{\Sigma}^{-1} (\mathbf{x}_i - \mathbf{x}_j) \right\}.
  \label{kernel}
\end{equation}

\noindent{The} bandwidth matrix $\mathbf{\Sigma}$ is a $2 \times 2$ 
positive definite covariance matrix:
%
\begin{equation}
    \mathbf{\Sigma} = \begin{pmatrix}
        \sigma_{1}^2 & \rho\sigma_{1}\sigma_{2} \\
        \rho\sigma_{1}\sigma_{2} & \sigma_{2}^2
    \end{pmatrix},
    \label{bandmat}
\end{equation}

\noindent{where} $\sigma_{1}, \sigma_{2} > 0$ 
is the scale parameter for each dimension and $-1 < \rho < 1$
is the correlation parameter.
Note that the assumption of a common kernel function across ecological states
simplifies the estimator (Eq. \ref{kernel.est2}) as follows:
%
\begin{equation}
  g_{its} = \frac{\sum_{j = 1}^I K_{\boldsymbol{\Sigma}}(\mathbf{x}_i, \mathbf{x}_j)\bm{1}(z_{jt} = s)}{\sum_{j = 1}^I K_{\boldsymbol{\Sigma}}(\mathbf{x}_i, \mathbf{x}_j)}.
  \label{kernel.est}
\end{equation}

With the kernel estimator of this form, $g_{its}$ becomes higher at site $i$ as the 
number of sites that are occupied by state $s$ at time $t$ increases.
Although all the sites occupied by $s$ at time $t$ contributes to $g_{its}$,
the weight decreases as a function of the distance from site $i$:
the scale of this distance dependence is controlled by
the bandwidth matrix $\mathbf{\Sigma}$. 
We note that $\mathbf{\Sigma}$ accounts for an anisotropic 
spatial dependence (i.e., direction dependence) of the local community structure:
the spatial dependence is isotropic only when $\sigma_{1} = \sigma_{2}$ and $\rho = 0$.

As long as there is no significant sources of observation error other than the 
resampling error we considered above,
the estimated vector $(g_{it1},\dots,g_{itS})$ is interpreted naturally as a representation of
the composition of ecological states \textit{surrounding} site $i$ at time $t$.
As in \cite{Fukaya2013}, $g_{its}$ is a derived parameter
obtained as a function of latent occupancy state \{$z_{it}$\},
whereas the function defining $g_{its}$ now depends on
an unknown bandwidth matrix with parameters to be estimated.
We also note that when the bandwidth of the kernel is sufficiently large,
$g_{its}$ will no longer vary over sites within the quadrat.
The result is a spatially homogeneous situation that is equivalent to the 
model considered by \cite{Fukaya2013}.

\subsection*{Estimation of states and parameters}

In this model, elements of the transition probability matrix $\mathbf{P}$,
the resampling error rate $e$, the initial occupancy probability vector $\boldsymbol{\phi}$,
and the elements of the bandwidth matrix $\boldsymbol{\Sigma}$ are parameters
to be estimated.
The model also involves occupancy state $z_{it}$, error occurrence indicator $m_{itj}$,
and the local state composition $g_{its}$ (a derived quantity)
as latent state variables.
The model assumes that observation at a particular time point depend on
all the state variables at that time.
Such a class of dynamic models is called a factorial hidden Markov model
in the machine learning community, and for such a model it is known that
an exact likelihood inference is computationally intractable \citep{Ghahramani1997}.
We thus adopt here a Bayesian approach for parameter inference,
in which the joint posterior distribution
of parameters and latent state variables is obtained by using 
Markov chain Monte Carlo (MCMC) methods.
We specify vague priors for parameters for the observation and process models:
%
\begin{equation}
    e \sim \textrm{Beta}(1, 1),
    \label{error.rate}
\end{equation}
%
\begin{equation}
    (p_{1s}, \dots, p_{Ss}) \sim \textrm{Dirichlet}(1, \dots, 1), 
    \label{prior.P}
\end{equation}
%
\begin{equation}
    (\phi_{1}, \dots, \phi_{S}) 
        \sim \textrm{Dirichlet}(1, \dots, 1),
    \label{phi}
\end{equation}
\noindent{where} the diffuse Dirichlet prior (Eq. \ref{prior.P}) is specified
for each of column $s$ of the transition probability matrix.
Vague priors would also be specified
for elements of the bandwidth matrix, $\boldsymbol{\Sigma}$, as follows:
%
\begin{equation}
    \begin{aligned}
        \sigma_{1} &\sim \textrm{Uniform}(0, U)
    \end{aligned}
\end{equation}
\begin{equation}
    \begin{aligned}
        \sigma_{2} &\sim \textrm{Uniform}(0, U)
    \end{aligned}
\end{equation}
%
\begin{equation}
    \rho \sim \textrm{Uniform}(-1, 1)
    \label{prior.rho}
\end{equation}

\noindent{where} $U$ is a sufficiently large positive constant
defining the upper limit of the uniform prior.

The joint posterior distribution of these parameters and latent variables
(\{$z_{it}$\}, \{$m_{itn}$\}, \{$g_{its}$\}) can be
obtained by using the MCMC methods.
The model can be implemented using the BUGS software.
We have provided a model script in the Appendix
that can be used to fit the model using JAGS software \citep{jags}.

\section*{SIMULATION STUDY}

\subsection*{Materials and methods}

To examine the effect of an observation process
that is relevant to the local community structure on the inference
of community dynamics, we conducted a simulation study.
Using the system and observation models described above,
with fixed model parameters that were to be estimated,
we simulated a number of replicate datasets
that were obtained from the 
hierarchical data-generating process described in the previous section.
Simulations and analyses described in what follows were
conducted using R (versions 3.1.0 to 3.2.5).

Three models were fitted to these simulated data sets.
The first was a classical, naive estimator for transition probabilities,
defined as $\hat{p}_{jk}=n_{jk}/\sum_l n_{lk}$, 
where $n_{jk}$ is the number of sites that were in state $k$ at
a certain point in time and in state $j$ at the following time
\citep{Spencer2005}.
This model does not account for any type of observation error.
The second was a ``non-spatial'' multistate dynamic site occupancy model
proposed by \cite{Fukaya2013},
which does account for resampling errors.
The model assumes that the distribution of the observed state,
conditional on the occurrence of the resampling error, is
proportional to the relative dominance of each state in the quadrat
that is homogeneous in space.
Finally, the ``spatial'' multistate dynamic site occupancy model
described in the previous section was also fitted to simulated datasets.
Note that this model is equivalent to the data-generating model
and was thus expected to perform the best among the three models.

To generate replicated datasets,
we set the number of sites $I=225$, the length of time series $T=5$,
and the number of ecological states $S=5$.
The sites were assumed to be aligned on a $15 \times 15$ grid,
and their coordinates were represented as 
$\mathbf{x}_1=(1,1), \mathbf{x}_2=(2,1), \dots, \mathbf{x}_{225}=(15,15)$.
We used a transition probability matrix that was randomly drawn from a diffuse
Dirichlet distribution
$(p_{1s}, \dots, p_{Ss}) \sim \textrm{Dirichlet}(1, \dots, 1)$.
For the observation process, we assumed an isotropic bandwidth kernel 
by setting $\sigma_1=\sigma_2=1$ and $\rho=0$.
With these settings, we generated 96 replicate datasets 
for each of 6 levels of error rate ($e=0, 0.15, 0.3, 0.45, 0.6$, $0.75$).

We considered two scenarios for the manner of data collection.
The first scenario considered a situation of limited information where
no replicated observations were obtained; that is, we set $N(i,t)=1$ for all $i$ and $t$.
The second scenario simulated a more ideal situation where
three replicated observations were conducted for each survey; that is,
we set $N(i,t)=3$ for all $i$ and $t$.

Among the three models and two sampling scenarios described above, 
the performances of the models were evaluated based on the mean square error (MSE),
square of bias ($\textrm{Bias}^2$) and variance (Var) of the point estimate of each parameter.
For transition probabilities, the above criteria were
averaged over all elements in the probability matrix
to evaluate the magnitude of the estimation errors
in a sequence of transition probabilities.
Because MSE measures the mean square difference of estimates from the true 
parameter value, it may serve as a comprehensive criterion for the 
quality of estimators.
The MSE can be decomposed into bias and variance,
as the three are linked by the relationship 
$\textrm{MSE}(\hat{\theta}) = \textrm{Bias}^2(\hat{\theta}) + \textrm{Var}(\hat{\theta})$.
Thus, we favor a model with a lower MSE, and such a model may even be better
if it provides less biased estimates.

For the naive model, data were simply aggregated to $n_{jk}$
to obtain estimates of transition probabilities.
Because this model does not accommodate replicated sampling,
it was not applied to data from the three-replicate scenario.
For the non-spatial and spatial model, parameters were estimated using
the Bayesian inference framework.
For each analysis of replicated datasets,
we specified a vague prior for each parameter 
(Eqs. \ref{error.rate}--\ref{prior.rho})
to sample from the posterior distribution using the 
JAGS software \citep[versions 3.4.0 to 4.2.0;][]{jags},
from three independent chains of 3000 iterations after a burn-in of 3000
iterations thinning at intervals of 3.
The posterior mode was used for point estimates of univariate parameters
($e$ and $\sigma$), whereas the spatial median was used
for transition probabilities to ensure that the sum of each column of the probability
matrix was 1.

Convergence of the posterior was checked for each parameter 
with the Gelman-Rubin diagnostic $\hat{R}$.
In the spatial model, about 0.03\% of the estimates of $p_{jk}$, 
9\% of $e$ (most of which came from small $e$
situations, where estimation of $\sigma$ was difficult; see the next subsection),
and 12\% of $\sigma$ (same as $e$) had $\hat{R} > 1.2$.
In the non-spatial model, about 0.3\% of $p_{jk}$ and 5\% of $e$ had $\hat{R} > 1.2$. 
Although these $\hat{R}$ values may not be appropriate in formal model inference
\citep[e.g.,][]{Gelman2014},
we considered our results to be acceptable for examination of the 
average performance of the model using the simulated dataset.
In practice, the $\hat{R}$ values could be improved by much longer MCMC runs,
which can be very time-consuming in the context of simulation but
would not be prohibitive for analysis of one or a small number of datasets.

\subsection*{Results}

\begin{figure*}[htb]
\begin{center}
  \includegraphics[width=3.75in]{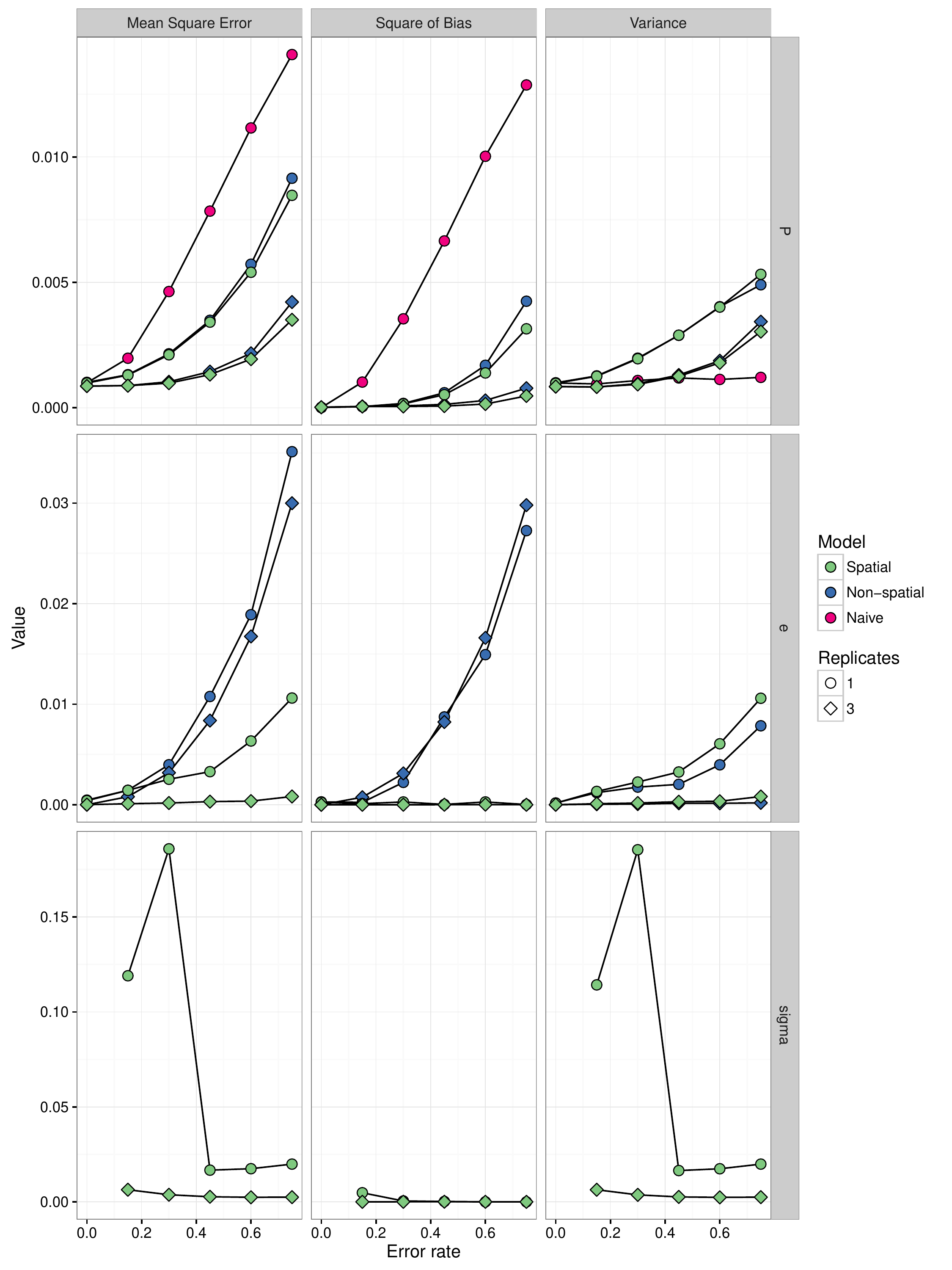}
\end{center}
  \caption{
    Results of simulation study.
    Mean square error (left column), square of bias (middle column), and variance (right column)
    of parameter estimates obtained from 96 replicated datasets using three different models
    (spatial model, non-spatial model, and naive estimator) are shown.
    Upper, middle, and lower panels indicate results for 
    the transition probability matrix $\hat{\mathbf{P}}$,
    resampling error rate $\hat{e}$ and the bandwidth scale parameter $\hat{\sigma}$, respectively.
    Results of $\hat{\sigma}$ for $e = 0$ in the spatial model are not shown
    because of a convergence issue (see text for details).
  }
  \label{fig:res_sim}
\end{figure*}

The results of the simulation study are summarized in Fig. \ref{fig:res_sim}.
We first note that when the error rate is small,
the posterior distribution for the bandwidth parameter $\sigma$ 
in the spatial model sometimes failed to converge.
In such cases, the posterior distribution of $\sigma$ was typically distributed
over a wide range, a reflection of the flat specified prior distribution,
the result being spatially homogeneous
estimates of the local state composition $g_{its}$.
This result suggests an identification issue for $\sigma$ in these settings,
an issue that appears to become worse when replicated samples 
are lacking (Fig. \ref{fig:res_sim}).
In the spatial model, $\sigma$ determines
the probability distribution of data \textit{conditional on} 
the occurrence of resampling error.
Hence, $\sigma$ is not satisfactorily determined by data when
resampling errors occur infrequently.
Even when the posterior sampling for $\sigma$ was unsuccessful,
however, the convergence of other parameters 
in the spatial model ($\mathbf{P}$ and $e$) was typically achieved.
In these cases, estimates of transition probabilities were
often very similar in the spatial model and the non-spatial model.
This similarity reflects the fact that the estimated local state composition $g_{its}$
was spatially homogeneous, the result being a spatial model that was virtually 
the same as the non-spatial model, in which a spatially homogeneous
state composition was assumed implicitly.

For this reason, we excluded estimates
from the calculation of MSE, bias, and variance of $\hat{\sigma}$,
if the difference between the point estimate and the true value of $\sigma$
was more than 10 (Fig. \ref{fig:res_sim}, lower panels).
We omitted the results of MSE, square bias, and 
the variance of $\hat{\sigma}$ for $e = 0$
because this procedure excluded almost all estimates for $e = 0$.
We also note that in the spatial model with $e=0$ and three replicated observations,
there was an estimate of $e$ that was exceptionally large.
We also excluded that estimate from the calculation of MSE, bias, and variance of $\hat{e}$.

Not surprisingly, overall, the bias was the smallest in the spatial model
compared to other models (Fig. \ref{fig:res_sim}).
When comparisons were made among the results of the single-replicate case,
estimates of transition probabilities from the naive estimator were the least variable
(Fig. \ref{fig:res_sim}, upper panels).
However, the estimator became profoundly biased as the error rate increased,
the result being the worst MSE of the transition probabilities among the examined models.
The bias and variance of the transition probability estimates 
in the non-spatial and spatial model increased with the error rate.
The bias of the transition probabilities was considerably smaller in these models
compared to the naive model, resulting in the better MSE as the error rate increased.
Although the difference between the non-spatial and spatial models
in the estimation of transition probabilities is not so large, 
the bias and MSE of the spatial model was smaller than that of the non-spatial model
when the error rate was high (i.e., $e \geq 0.45$).
Replicated observations improved the estimation of
transition probabilities in the spatial and non-spatial model. 
Under these conditions, the performance of the non-spatial model was
even better than that of the spatial model with a single replicate.

The MSE of $\hat{e}$ tended to be higher for the non-spatial model than for the spatial model
(Fig. \ref{fig:res_sim}, middle panels)
because the bias of the non-spatial model increased faster than that of the spatial model,
although the variance of $\hat{e}$ was in general smaller in the non-spatial model
than in the spatial model.
In the non-spatial model, replicated observations did not mitigate the MSE and bias of $\hat{e}$.
In the spatial model, the MSE and bias of $\hat{\sigma}$ tended to be higher when the error
rate was low (Fig. \ref{fig:res_sim}, lower panels), where, as noted above,
the estimation of $\sigma$ may be difficult.

In summary, these results suggest that, when the local community structure
affects observations, both the naive estimator and 
the non-spatial multistate dynamic site occupancy model may provide worse estimates
than the spatial model.
When the error rate is low, the bandwidth parameter of the spatial model may 
not be satisfactorily determined by data, but the model may still perform as well as 
the non-spatial model by estimating the spatially homogeneous relative state composition.
As the error rate increased, the spatial model on average outperformed the non-spatial model.

\section*{APPLICATION TO REAL DATA}

\subsection*{Materials and methods}

We applied the proposed spatial model,
in addition to the non-spatial model \citep{Fukaya2013}
and the naive estimator for transition probabilities \citep{Spencer2005},
to an intertidal sessile community dataset
collected at a total of 25 permanent quadrats within 5 shores 
--- Mochirippu (MC), Mabiro (MB), Aikappu (AP), Monshizu (MZ) and Nikomanai (NN)---
located on the Pacific coast of eastern Hokkaido, Japan
(detailed descriptions of the study sites and biogeographic features of the area
can be found in 
\citealp{Okuda2004}, \citealp{Nakaoka2006} and \citealp{Fukaya2014}).
Each quadrat was established on rock walls at semi-exposed locations
and was 50 cm wide by 100 cm high.
The vertical mid-point corresponded to the mean tidal level.
In each quadrat, 200 fixed observation points (i.e., sites)
were aligned on 20 horizontal rows by 10 vertical columns.
The sites were located at intervals of 5 cm.
Site occupancy dynamics of ecological states were assessed once per year from 2002 to 2011,
during a spring low tide in the summer (from roughly July to August).

\begin{table*}[htb]
  \centering
  \caption{
    Parameter estimates (transition probability matrix $\mathbf{P}$,
    resampling error probability $e$, and bandwidth matrix $\boldsymbol{\Sigma}$)
    obtained by fitting each of the three different models
    (naive estimator, non-spatial model, and spatial model)
    to the intertidal sessile community dataset of a particular quadrat (MC2).
    Posterior medians are reported for the spatial and non-spatial models,
    for which posterior sampling was conducted using MCMC.
    The rows and columns of the transition probability matrix correspond to:
    (1) free space, (2) the barnacle \textit{Chthamalus dalli}, 
    (3) the red alga \textit{Gloiopeltis furcata}, 
    (4) the articulated calcareous red alga \textit{Corallina pilulifera}, and
    (5) other organisms.
    }
  {\tiny
  \begin{tabular}{lccc} \toprule
     & \multicolumn{3}{c}{Model} \\
    Parameter & Naive & Non-spatial & Spatial \\
    \midrule
    $\mathbf{P}$ & 
      $\begin{bmatrix}
        0.560 & 0.328 & 0.375 & 0.051 & 0.164 \\
        0.085 & 0.267 & 0.100 & 0.020 & 0.082 \\
        0.320 & 0.339 & 0.446 & 0.020 & 0.143 \\
        0.006 & 0.013 & 0.003 & 0.471 & 0.156 \\
        0.028 & 0.053 & 0.076 & 0.438 & 0.455
      \end{bmatrix}$ &
      $\begin{bmatrix}
        0.670 & 0.221 & 0.307 & 0.013 & 0.028 \\
        0.064 & 0.416 & 0.100 & 0.014 & 0.016 \\
        0.258 & 0.328 & 0.585 & 0.011 & 0.020 \\
        0.003 & 0.005 & 0.002 & 0.501 & 0.196 \\
        0.005 & 0.029 & 0.006 & 0.461 & 0.740
      \end{bmatrix}$ & 
      $\begin{bmatrix}
        0.772 & 0.160 & 0.208 & 0.015 & 0.040 \\
        0.053 & 0.574 & 0.060 & 0.016 & 0.029 \\
        0.165 & 0.239 & 0.713 & 0.015 & 0.037 \\
        0.003 & 0.008 & 0.002 & 0.684 & 0.098 \\
        0.007 & 0.020 & 0.017 & 0.270 & 0.796
      \end{bmatrix}$ \\
    $e$ & & $0.239$ & $0.715$  \\
    $\mathbf{\Sigma}$ & & & $\mbox{diag}(0.644, 1.278)$ \\
    \bottomrule
  \end{tabular}
  }
  \label{table:estimates}
\end{table*}

Models were fitted to estimate transition probabilities for each quadrat.
In the study sites, we recognized the following seven dominant ecological states:
(1) free space, (2) the barnacle \textit{Chthamalus dalli}, 
(3) the brown alga \textit{Analipus japonicus}, (4) the red alga \textit{Gloiopeltis furcata},
(5) the red alga \textit{Chondrus yendoi}, (6) the articulated calcareous red alga
\textit{Corallina pilulifera}, and (7) other organisms.
However, states 1--6 were not necessarily common in each quadrat.
Thus, for each quadrat, we allocated some of these regionally dominant states 
to state 7 (other organisms) when they were relatively rare 
(when the total number of their occurrences during the 10 observation years was $< 50$)
in that quadrat.
As a result, the number of states varied from 4 to 7 among quadrats.

For each quadrat, models were fitted to data in a fashion analogous to
that used in the simulation study described in the previous section.
To fit the spatial model, we restricted the bandwidth matrix to
$\boldsymbol{\Sigma} = \textrm{diag}(\sigma_1^2, \sigma_2^2)$,
where $\sigma_1$ and $\sigma_2$ were the scale parameters along the vertical and
horizontal directions, respectively.
For spatial and non-spatial models, 
we first conducted MCMC sampling for each plot with three independent chains of
3000 iterations after a burn-in of 3000 iterations, thinning at intervals of 3.
For some plots, however, we needed to have longer iterations and more chains
to obtain converged posterior samples.
The convergence of parameters of interest was confirmed if
the Gelman-Rubin diagnostic $\hat{R}$ was less than 1.1.
We note, however, that for MB4, AP4 and AP5, we could not obtain converged
posterior samples of $\sigma_2$, probably because the spatial dependency
along the horizontal axis was lacking or because there were insufficient data to estimate it.

As derived parameters, we also obtained estimates of two quantities that 
inform some dynamical aspects of communities:
the mean turnover time and the damping ratio \citep{Hill2004}.
The former quantifies the average turnover time of a site
randomly selected from the stationary community.
It is calculated as $\sum_{s=1}^{S}w_s (1 - p_{ss})^{-1}$
\noindent{where} $w_s$ is the equilibrium relative dominance of state $s$
obtained as the normalized dominant right eigenvector of the 
transition probability matrix $\mathbf{P}$.
The latter quantifies the lower bound of the rate of
convergence to equilibrium. It is equated to
$1/|\lambda_2|$ where $\lambda_2$ is the 
second-largest eigenvalue of $\mathbf{P}$.

\subsection*{Results}

\begin{figure*}[htb]
\begin{center}
  \includegraphics[width=4.5in]{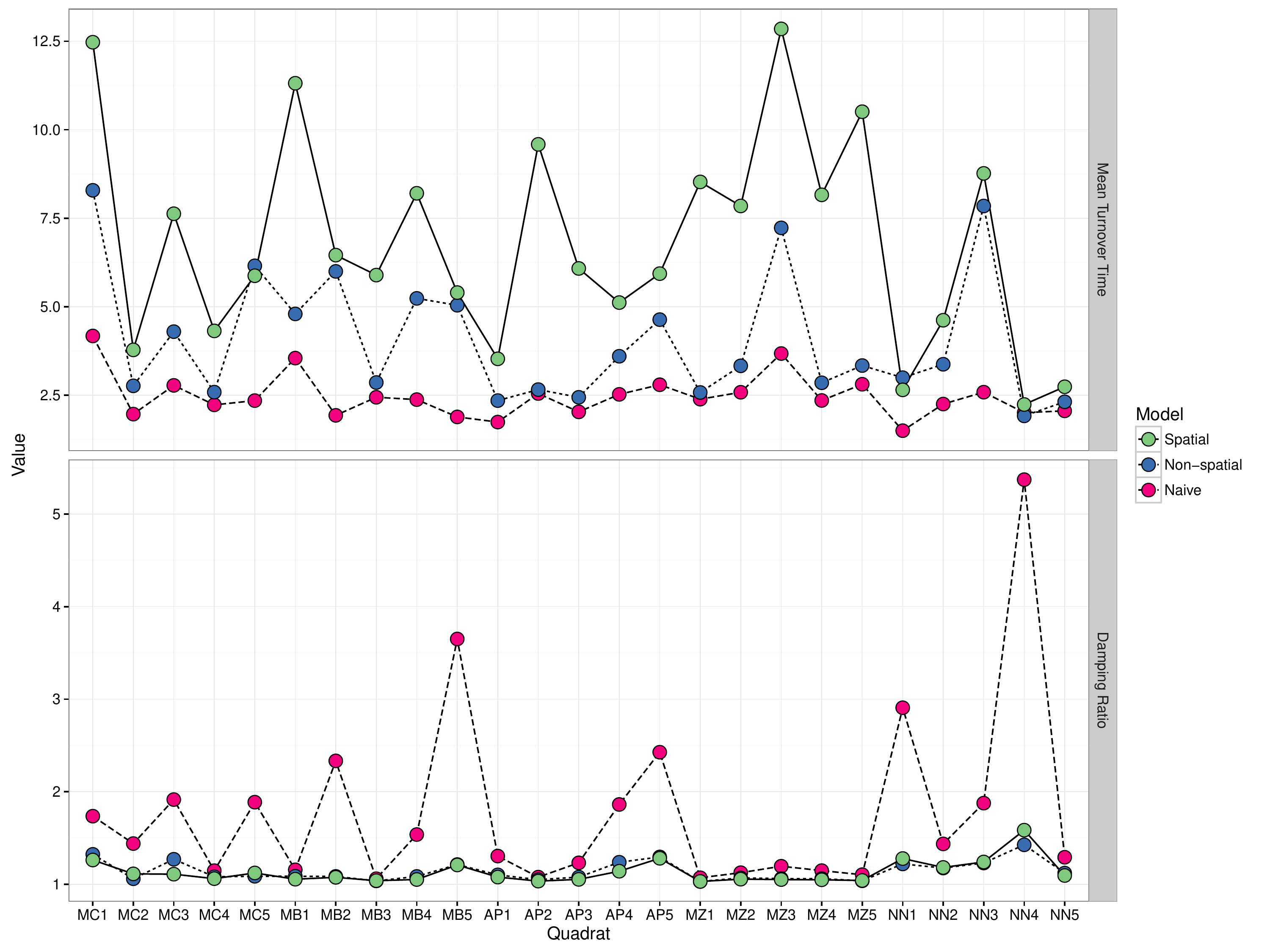}
\end{center}
  \caption{
    Mean turnover time (upper panel) and damping ratio (lower panel)
    estimated using three models (spatial model, non-spatial model, and naive estimator)
    for each quadrat.
  }
  \label{fig:comm_prop}
\end{figure*}

As an example, parameter estimates obtained at a particular quadrat (MC2)
for each model are shown in Table \ref{table:estimates}.
The estimated diagonal elements of the transition probability matrix
(i.e., the probability of persistence of the occupied state)
tended to be higher for the non-spatial model
than those obtained with the naive estimator.
These estimated probabilities were even higher for the spatial model.
This tendency was generally found in other quadrats
and reflects a consistent pattern of 
estimated properties of community dynamics (Fig. \ref{fig:comm_prop}).
On the one hand, the estimated mean turnover time tended to be longest in the spatial model,
intermediate in the non-spatial model, and shortest in the naive method.
On the other hand, the damping ratio was estimated to be fairly high
for the naive method, whereas it was lower for the spatial and non-spatial models.
These results suggest that ignoring the resampling error
and the effect of local community structure on observations
may lead to an overestimation of the ``velocity'' of community dynamics
because of an underestimation of persistence probabilities.

\begin{figure*}
  {\centering
    \includegraphics[bb=0 0 1152 432,width=16.5cm]{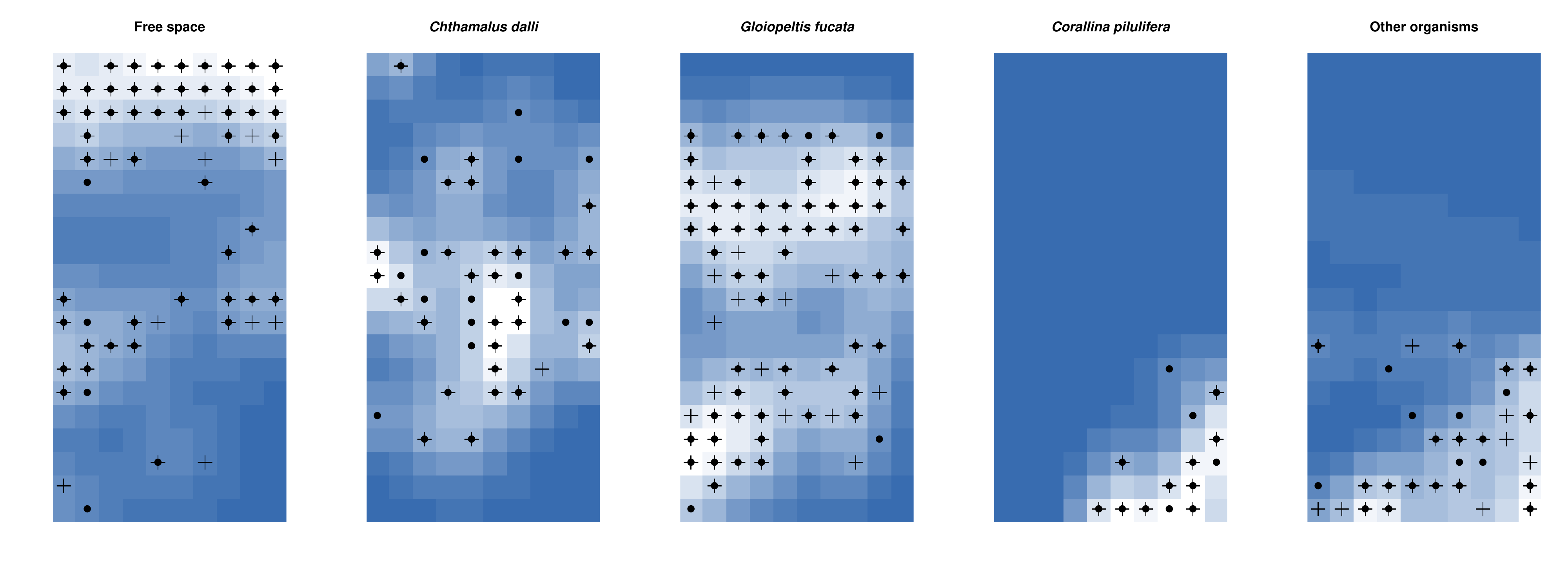}
  }
  \caption{
    Observed data, occupancy, and local state dominance for
    a particular quadrat (MC2) and point in time estimated using the spatial model,
    mapped on the spatial coordinates of each observation point 
    (20 horizontal rows $\times$ 10 vertical columns).
    Filled circles ($\bullet$) and crosses ($+$) represent data 
    (the state observed at each observation point)
    and estimated site occupancy state (the state that had the highest posterior probability
    of occupancy at each observation point), respectively.
    Note that for each site, both filled circles and crosses appear in only 
    one of the five panels, which represent different ecological states.
    The estimated local state dominance, $g_{its}$, is represented by
    colored squares. A brighter square indicates greater dominance of the state at that location.
  }
  \label{fig:est_f}
\end{figure*}

Interestingly, in the spatial model, the scale of bandwidth for
the vertical axis was estimated to be approximately half 
that for the horizontal axis (Table \ref{table:estimates}).
This result coincides with the known observation that
rocky intertidal communities may be more variable
along the vertical than the horizontal direction
because environments tend to vary more rapidly 
along the vertical axis \citep{Benedetti-Cecchi2001, Valdivia2011}.
The estimated resampling error rate tended to be higher
in the spatial model than in the non-spatial model (Table \ref{table:estimates}).
Observations on these parameters noted here 
were also generally found in other quadrats.
We show in Fig. \ref{fig:est_f} a snapshot of
the estimated local community composition 
for a particular quadrat (MC2) and point in time.

\section*{DISCUSSION}

In this study, we proposed a multistate dynamic site occupancy model
for aggregated sessile communities.
The model accounts for the resampling error
that is associated with local community structure.
The model assumes underlying surfaces of local dominance for each of the
possible ecological states
that determines the distribution of data that are conditional on the 
occurrence of the resampling error.
The surface is estimated nonparametrically by using a multivariate kernel 
smoothing method, in which 
the local species dominance at an arbitrary location is obtained 
as a function of the underlying occupancy of states across all sites.

In ecological studies, consideration of spatial effects has been
thought to be fundamental \citep{Legendre1993,Lichstein2002,Royle2007}.
It has also been widely recognized that accounting for observation processes 
in data collection and analyses is important because ignoring these processes may lead to
biased estimates and thus misleading conclusions \citep{Williams2002,Kery2012,Kery2015}.
These two considerations have been incorporated into the
dynamic site occupancy modeling framework to study metapopulation dynamics
while accounting for imperfect detection 
\citep[e.g.,][]{Bled2011, Risk2011, Yackulic2012}.
As far as we know, however, such spatial dynamic site occupancy models
have been applied only to population dynamics of single species (but see \cite{Yackulic2014}).
The spatial multistate dynamic site occupancy model we developed here
would be widely applicable to studying the dynamics of sessile communities
in various systems, including corals, mussels and terrestrial plants,
for which precise resampling of observation points may be
difficult in the field for practical reasons \citep{CCD2011}.

For spatially referenced site occupancy data, dynamic site occupancy models with
autologistic components have been proposed
\citep[; see also \citealt{Broms2016}]{Bled2011, Bled2013, Yackulic2012, Yackulic2014, Eaton2014}.
In these models, the adjacency of sites must be defined {\it a priori},
because probabilities of occupancy, local colonization and extinction are assumed to
depend on the occupancy states of neighboring sites.
It has also been typical in the application of these models that sites are aligned on grids.
However, as has been done in this study, another approach for
modeling spatial effects is to use kernel-weighting functions.
Several dynamic site occupancy models that were recently developed based on the
metapopulation theory use this approach to account for 
spatial dependency of local colonization and extinction probabilities
\citep{Risk2011, Heard2013, Sutherland2014, Chandler2015}.
Because the scale of correlation is determined by the data,
this modeling approach does not require {\it a priori} determination of site adjacency.
The observation sites may also not need to be aligned in a reticular pattern.
A related weighting function approach has also been used to identify
the spatial scale at which landscape variables affect abundance \citep{Chandler2016}.
We note, however, that a possible difficulty in using kernel weighting may be
its associated computational burden.
Evaluating the kernel weights for every combination of two sites
may be time-consuming, and the amount of computation time required 
increases very rapidly with the number of sites being considered.

The multivariate Gaussian kernel used in the proposed model 
can account for an anisotropic spatial correlation of community structure.
This characteristic of the multivariate Gaussian kernel
can be useful when a known, or even unknown, environmental gradient exists 
in the quadrat and affects the distribution of sessile species.
Although other types of kernels may be used to model the spatial structure,
the multivariate Gaussian kernel allows a straightforward implementation of 
anisotropic correlation by specifying a bandwidth matrix 
with scale and correlation parameters.
However, results of our simulation study suggest that, in the proposed model,
bandwidth parameters may be difficult to estimate when the resampling error rate
is quite low.
In such cases, the proposed spatial model would not be a reasonable option
for the inference of transition probabilities.
In this regard, adoption of a robust design is recommended because 
obtaining replicated samples may mitigate this problem (Fig. \ref{fig:res_sim}).
We note that dynamic occupancy models in general do not require equal numbers of
replications for every site and time.

Results from a dataset of intertidal communities on the Pacific coast of
eastern Hokkaido highlight the fact that different models may produce
considerably different estimates of transition probabilities
as well as of the properties of community dynamics, which are obtained from
the transition probability matrix 
(Table \ref{table:estimates}, Fig. \ref{fig:comm_prop}).
We have shown that, compared to the non-spatial model \citep{Fukaya2013},
the spatial model tends to yield a higher persistence probability
of each state, longer mean turnover time and lower damping ratio.
These differences were even greater when the spatial model was compared to the naive model.
In the assessment of their intertidal sessile community data from California,
\cite{CCD2011} also found that the estimated damping ratio was higher when
resampling errors were not corrected.
These results have implications for the development of management plans
that are based on Markovian models.
They suggest that ignoring the existence of resampling errors, as well as
local community structure, may lead to overestimation of the recovery rate of
ecological communities from natural or anthropogenic disturbance,
the result being poorly informed management strategies.

The hierarchical formulation of the model allows us to readily extend the proposed model,
at least conceptually, to add more ecological realism \citep{Royle2008,Kery2012,Kery2015}.
For example, if some site- or time-specific environmental covariates are available,
they may be incorporated into the model to take account of variations in
transition probabilities and error rates.
A possible extension of the model that may have a particular ecological significance is
to link estimated local state dominance $g_{its}$ values to transition probabilities.
Because interactions between sessile organisms are limited to neighboring individuals,
the rate of local colonization, persistence, and replacement are expected to depend
on surrounding local community structure \citep{Wootton2001,Kawai2006}.
Although the dependency of transition probabilities on species 
frequency has been considered in previous studies about Markovian community dynamics
(i.e., nonlinear Markov community model; \citealt{Spencer2008,Tanner2009}),
only the connection between transition probabilities and the overall (i.e., global)
species frequency has been explored.
However, by using the underlying occupancy component explicitly,
dynamic site occupancy models can model a
relationship between local occupancy density and the rates of
colonization and/or extinction \citep{Bled2011, Risk2011, Yackulic2012},
although such dependency has rarely been modeled in a multispecies context
(but see \cite{Yackulic2014} who explored intra- and inter-specific effects
on site occupancy dynamics of two owl species).
Because the proposed model explicitly involves local community structure as a model component, 
it will naturally provide a basis for inferring complex nonlinear ecological dynamics.

\section*{ACKNOWLEDGEMENTS}
We thank I. K. Shimatani and S. Eguchi for their helpful comments and discussion.
For providing access to laboratory facilities, we are grateful to the staff and
students of the Akkeshi Marine Stations of Hokkaido University.
We acknowledge many researchers and students who helped with our fieldwork.
This research was supported by an allocation of computing resources of 
the SGI ICE X and SGI UV 2000 supercomputers from the Institute of Statistical Mathematics.
Funding was provided by the Japan Society for the Promotion of Science
(Grants-in-Aid for Scientific Research No 15K18617
and Grant-in-Aid for JSPS Fellows No 16J07614 to KF,
Grants-in-Aid for Scientific Research Nos
20570012, 24570012 and 15K07208 to TN, 
and Nos 14340242, 18201043 and 21241055 to MN).

\clearpage
\newpage

\appendix

\begin{@twocolumnfalse}

  \section{JAGS model code}
  \begin{verbatim}
model {
    ## Observation model
    for (i in 1:I) {
        for (t in 1:T) {
            for (n in 1:N[i, t]) {
                # m: indicator for resampling error, 1 represents resampling error
                m[i, t, n] ~ dbern(e)
                for (s in 1:S) {
                    # q: observation probabilities (conditional on m)
                    q[i, t, n, s] <- ((1 - m[i, t, n]) * equals(z[i, t], s)
                        + m[i, t, n] * g[i, t, s])
                }
                # y: observed data
                y[i, t, n] ~ dcat(q[i, t, n, ])
            }
        }
    }

    ## Process model
    for (i in 1:I) {
        # z: occupancy state
        z[i, 1] ~ dcat(phi[])
        for (t in 2:T) {
            z[i, t] ~ dcat(p[, z[i, t - 1]])
        }
    }

    ## Kernel estimator
    # Sigma: bandwidth matrix
    Sigma[1, 1] <- sigma1^2
    Sigma[1, 2] <- rho * sigma1 * sigma2
    Sigma[2, 1] <- rho * sigma1 * sigma2
    Sigma[2, 2] <- sigma2^2

    # V: inverse of the bandwidth matrix
    V <- inverse(Sigma)

    for (i in 1:I) {
        for (j in 1:I) {
            # K: weight matrix
            K[i, j] <- exp(- vd[i, j, ] %*% V %*% vd[i, j, ] / 2)
        }
    }

    for (i in 1:(I - 1)) {
        # dS: Diagonal matrix of the inverse sum of weights
        dS[i, i] <- 1 / sum(K[i, ])
        for (j in (i + 1):I) {
            dS[i, j] <- 0
            dS[j, i] <- 0
        }
    }
    dS[I, I] <- 1 / sum(K[I, ])

    for (t in 1:T) {
        # mz: array of indicator of occupancy state
        for (s in 1:S) {
            mz[1:I, t, s] <- equals(z[, t], s)
        }
        # g: relative state dominance
        g[1:I, t, 1:S] <- dS %*% K %*% mz[, t, ]
    }

    ## Priors
    # e: resampling error rate
    e ~ dbeta(1, 1)

    # P: transition probabilities
    for (s in 1:S) {
        p[1:S, s] ~ ddirch(alpha)
    }

    # phi: initial occupancy probabilities
    phi[1:S] ~ ddirch(alpha)

    # sigma: scale parameter of the Gaussian kernel
    sigma1 ~ dunif(0, U)
    sigma2 ~ dunif(0, U)

    # rho: correlation parameter of the Gaussian kernel
    rho ~ dunif(-1, 1)
}
  \end{verbatim}

\end{@twocolumnfalse}


\end{document}